\documentclass[usenatbib]{mnras}
\usepackage{lscape}
\usepackage{amsmath,amssymb}
\usepackage{mathrsfs}	
\usepackage{multicol}
\usepackage[dvipdfmx]{graphicx}

\catcode`\@=11
\def\gta{\ifmmode{\,\mathrel{\mathpalette\@versim>\,}}
    \else{$\,\mathrel{\mathpalette\@versim>}\,$}\fi}
\def\lta{\ifmmode{\,\mathrel{\mathpalette\@versim<\,}}
    \else{$\,\mathrel{\mathpalette\@versim<}\,$}\fi}
\def\@versim#1#2{\lower 2.9truept \vbox{\baselineskip 0pt \lineskip
    0.5truept \ialign{$\m@th#1\hfil##\hfil$\crcr#2\crcr\sim\crcr}}}
\catcode`\@=12  

{\newif\ifnotend
\notendtrue
\def\veclist{ABCDEFGHIJKLMNOPQRSTUVWXYZabcdefghijklmnopqrstuvwxyz.}
\def\top#1#2.{#1}
\def\tail#1#2.{#2.}
\loop\expandafter\xdef\csname v\expandafter\top\veclist\endcsname%
{{\noexpand\bf\expandafter\top\veclist}}
\edef\veclist{\expandafter\tail\veclist}
\if\veclist.\notendfalse\fi\ifnotend\repeat}
\def\d{{\rm d}}\def\p{\partial}
\def\fracj#1#2{{\textstyle{#1\over#2}}}

\newcommand{\bJ}{{\bf J}}
\newcommand{\bx}{{\bf x}}
\newcommand{\bv}{{\bf v}}
\newcommand{\bL}{{\bf L}}

\newcommand{\fJ}{f({\bf J})}

\newcommand{\kJ}{k({\bf J})}
\newcommand{\Jf}{J_0}
\newcommand{\Jr}{J_r}
\newcommand{\Jphi}{J_{\phi}}
\newcommand{\Jz}{J_z}
\newcommand{\Ji}{J_i}
\newcommand{\tJphi}{\widetilde{J}_\phi}
\newcommand{\tJz}{\widetilde{J}_z}
\newcommand{\Lz}{L_z}

\newcommand{\etaphi}{\eta_{\phi}}
\newcommand{\etaz}{\eta_{z}}

\newcommand{\kphi}{k_{\phi}}
\newcommand{\kphieta}{k_{\phi} = \eta_{\phi}}
\newcommand{\kphixi}{k_{\phi} = \xi}

\newcommand{\Hphi}{\xi}  
\newcommand{\mIJ}{m_1}
\newcommand{\mIIJ}{m_2}

\newcommand{\PhiS}{\Phi_{\rm S}}

\newcommand{\DD}{\partial}
\newcommand{\dd}{\text{d}}

\newcommand{\Mi}{M_i}
\newcommand{\rc}{r_{\rm c}}

\newcommand{\rhoc}{\rho_{\rm c}}

\newcommand{\Ri}{R_i}
\newcommand{\zi}{z_i}
\newcommand{\sumiN}{\sum}
\renewcommand{\[}{\begin{equation}}
\renewcommand{\]}{\end{equation}}

\begin{document}

\date{Submitted, 21 July 2019}
\title[]{Flattened stellar systems based on distribution functions depending on actions}
{}
\author[]{Raffaele Pascale$^{1,2}$\thanks{E-mail: raffaele.pascale2@unibo.it}, James Binney$^{3}$ and Carlo Nipoti$^{1}$
\\ \\
$^{1}$Dipartimento di Fisica e Astronomia, Universit\`a di Bologna, via Piero Gobetti 93/2, I-40129 Bologna, Italy \\
$^{2}$Osservatorio di Astrofisica e Scienza dello Spazio, via Piero Gobetti 93/3, I-40129 Bologna, Italy \\
$^{3}$Rudolf Peierls Centre for Theoretical Physics, Clarendon Laboratory, Oxford OX1 3PU, United Kingdom}

\maketitle

\begin{abstract}
We address an issue that arises when self-consistently flattened dynamical stellar
systems are constructed by adopting a distribution function (DF) that depends
on the action integrals. The velocity distribution at points on the symmetry
axis is controlled by the the dependence of the DF on just one action, while
at points off the symmetry axis two actions are involved. Consequently, the physical
requirement that the velocity distribution evolves continuously in the
neighbourhood of the symmetry axis restricts the functional forms of
acceptable DFs. An algorithm for conforming to this restriction is presented
and used to construct a variety of flattened models.
\end{abstract}


\section{Introduction}
\label{sec:int}

Model stellar systems with known distribution functions (DFs) $f(\bx,\bv)$
are powerful tools for the interpretation of observations because such a
model predicts the outcome of any observation. For example, the system's
surface brightness can be obtained by integrating $f$ times the luminosity
per star over velocities and the line of sight; a map of any velocity moment
can be obtained by including an appropriate power of $\vv$ in the integral.
If the system's stars are resolved, the likelihood of the data given the
model can be computed regardless of how many phase-space coordinates have
been measured and with what precision.

In general it is essential to consider a stellar system to comprise several
components. For example, a globular cluster will contain a range of stellar
masses and include significant numbers of massive, dark, remnants. In
addition to these components, galaxies are thought to contain a population
of dark-matter particles that are distributed more extensively in phase space
than the stars, and they generally contain several stellar populations that
differ by age and/or chemical composition. 

A small number of model systems are known that have analytic expressions for
density $\rho(\bx)$ and potential $\Phi(\bx)$ and also analytic expressions
for the DF. All such systems are spherical and the only multi-component
models are  the rather specialised and complex two-component models described by
\cite{Ciotti1999}. A wider range of models can be obtained by considering systems
with DFs that are analytic functions of integrals of stellar motion but have
density and potential distributions that have to be obtained numerically.
Traditionally the integrals of stellar motion used as arguments of the DF
have been the energy $E=\frac12 v^2+\Phi(\bx)$ and the magnitude of the
angular momentum $\bL=\bx\times\bv$. However, the key to producing
multi-component and aspherical systems is to take the DF to be a specified
function $\fJ$ of the action integrals $\Ji$. Advantages of using actions
as  arguments of the DF include

\begin{itemize}
 \item The mass of any component is specified by the DF  before the system's
 density and potential have been determined.

 \item The addition of an extra component of mass $\Mi$ changes the density
 distributions of other components in an intuitive way.

 \item The self-consistently generated potential can be solved for by a stable
 and rapidly convergent iteration.
\end{itemize}

\noindent Early uses of DFs that depended exclusively on actions examined the
structure predicted by DF in assumed potentials \citep{Binney2010,Binney2012b}. 
The potential self-consistently generated by a DF $\fJ$ was first obtained by
\cite{Binney2014}, but only for one-component systems. \cite{PifflPenoyreB} 
solved for the potential self-consistently generated by a multi-component DF 
designed to model our Galaxy. \cite{Posti2015}, \cite{Williams2015} and \cite{Pascale2018}
explored spherical models of early-type galaxies that are defined by DFs $\fJ$ 
designed to model both dark haloes and the stellar content of elliptical and 
dwarf spheroidal galaxies.

In the case of a spherical potential, the actions comprise two components
$\Jphi$ and $\Jz$ of the angular momentum $\bL$, and the radial action
$\Jr$. $\Jphi$ is the component of $\bL$ about some chosen axis, and
$\Jz=L-|\Jphi|$ quantifies the inclination of the orbit with the respect to
the chosen axis. The radial action $\Jr=(2\pi)^{-1}\oint\dd r\,p_r$ quantifies
the amplitude of a star's radial oscillations.

If the part of $f$ that is an even function of $\Jphi$ depends on $\Jphi$
and $\Jz$ only through the combination $L=\Jz+|\Jphi|$, the system's
real-space structure will be spherical. It may, however, have net rotation
around the  axis that defines $\Jphi$: rotation around this axis is
encoded in the part of $f$ that is a odd function of $\Jphi$.
If the part of $f$ that is even in $\Jphi$ depends on $\Jphi$ and $\Jz$
other than through the combination $L$, the model will be aspherical. If
$f$ decreases with increasing $\Jz$ faster than it does with increasing
$|\Jphi|$, the model will be oblate.

In an oblate potential, the existence of actions is not guaranteed, but
numerical orbit integrations in plausible oblate potentials reveal that the
great majority of orbits are quasiperiodic \citep{BinneyTremaine2008}, which implies the existence of
three action integrals. For the majority of orbits these actions prove to be
minor modifications of the actions $\Ji$ familiar from the spherical case
\citep{JJBPJM16}. A
minority of `resonantly trapped' orbits have actions that are not simply
related to the spherical actions, although they can be computed using
first-order perturbation theory formulated in terms of the usual $\Ji$
\citep{Binney2016}. To a good approximation, the existence of trapped orbits
can be neglected when model building since  an ensemble of trapped 
orbits generate very similar predictions for most observables to an 
ensemble of untrapped orbits \citep{Binney2018}.

Subject to this proviso regarding the treatment of resonantly trapped orbits,
{\it any} non-negative function of three variables for which the integral
$\int\d^3\vJ\,\fJ$ through the positive octant of Cartesian space is
finite defines a stellar model of mass $M$, because, given such a function, 
one can normalise it such that $(2\pi)^{-3}=\int\d^3\bJ\,\fJ$ and can then 
solve for the potential $\Phi$ that satisfies

\[\label{eq:Poisson}
\nabla^2\Phi(\bx)=4\pi GM\int\d^3\bv\,f[\bJ(\bx,\bv)].
\]

This computation is rendered feasible by the St\"ackel Fudge
\citep{Binney2012b}, which, given any plausible axisymmetric potential
$\Phi(\bx)$, provides approximate formulae for $\bJ(\bx,\bv)$. Equation
(\ref{eq:Poisson}) can be solved in $\sim 5$ iterations, starting from
any plausible initial potential $\Phi_0$, by taking the potential $\Phi_{n+1}$ to
be that on the left of the equation with  $\Phi_n$ used in the
St\"ackel Fudge on the right \citep[][hereafter B14]{Binney2014}.

While any non-negative, normalisable function $\fJ$ defines a logically
possible model, B14 already noted that unless candidate functions $\fJ$ are
subjected to restrictions, the final model is liable to display physically
implausible structure near the origin and/or symmetry axis. Moreover,
\cite{PifflPenoyreB} discovered that the simplest DFs for our Galaxy's dark
halo predicted implausibly cusped velocity distributions.
The goal of this paper is to elucidate conditions on $\fJ$ that ensure
that it will generate a physically plausible model.

In Section~\ref{sec:ideas} we explain the physical origins of the
restrictions on $\fJ$ and suggest a way of satisfying them. In
Section~\ref{sec:example} we illustrate the effectiveness of our proposal by
presenting a variety models with and without implementation of our proposal.
In Section~\ref{sec:discuss} we ask why galaxies are restricted in the
distribution of stars in action space, and Section~\ref{sec:conclude} sums up.

\section{Restricting the DF}\label{sec:ideas}

\begin{figure*}
\centerline{\includegraphics[width=.4\hsize]{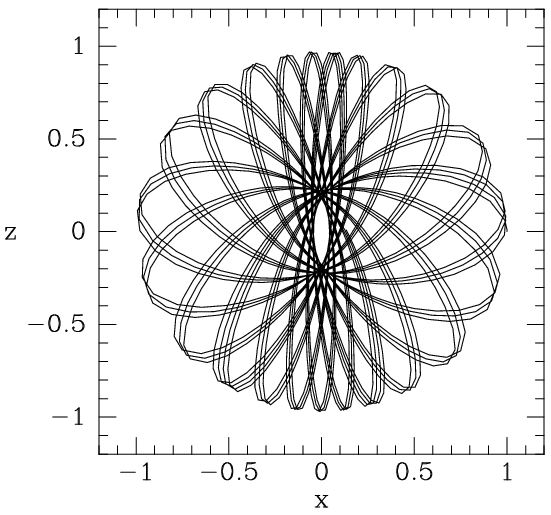}
\quad\includegraphics[width=.4\hsize]{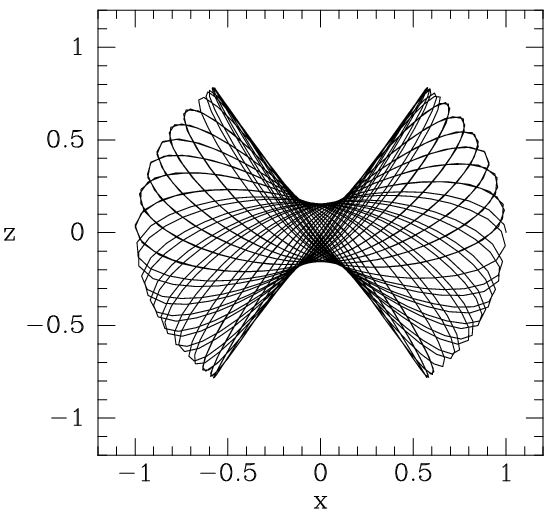}} \caption{Two orbits with
$\Jphi=0$ in a flattened potential. The loop orbit in the left panel avoids
the central section of the $z$ axis, while the box orbit on the right visits
no other part of the $z$ axis. Hence the system's kinematics parallel to the
$xy$ plane is controlled by box orbits in the central section of the $z$
axis, and by loop orbits outside this section. Here we show sections through
orbits that are rotationally symmetric about the $z$ axis.}\label{fig:one}
\end{figure*}

\subsection{Physical motivation}\label{sec:motiv}

Along the symmetry axis of an axisymmetric galaxy, the two directions that
run parallel to the equatorial plane are physically equivalent. If we call
these the $x$ and $y$ directions, it follows that at any point on the
symmetry axis, the distributions of $v_x$ and $v_y$ must be identical.

Only orbits with $\Jphi=0$ can reach the symmetry axis, so when considering
the velocity distribution at points on that axis we can confine attention to
the plane $\Jphi=0$ of action space, which has axes $\Jr$ and $\Jz$. The
orbits in this plane fall into two families: boxes (with small values of
$J_z$) and loops (with $J_z$ above a threshold), as illustrated
by Fig.~\ref{fig:one}. The loop orbits do not reach a central section
of the symmetry axis; this section is reached by box orbits, which do not
visit the part of the symmetry axis that is visited by loops.

\begin{figure}
\centerline{\includegraphics[width=.48\hsize]{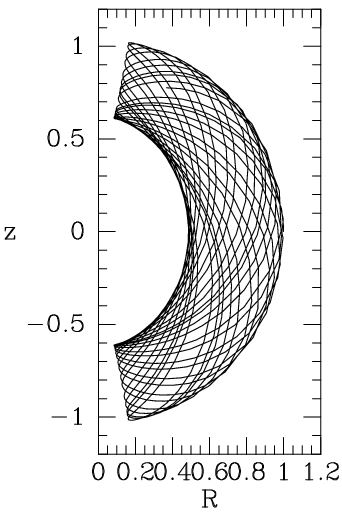}
\includegraphics[width=.48\hsize]{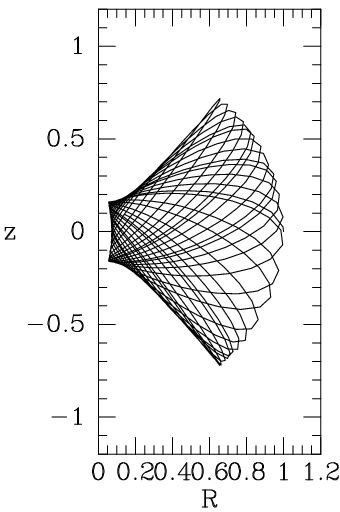}}
\caption{Two orbits with small but non-zero $J_\phi$. As $J_\phi\to0$, the
orbit on the left tends to a loop orbit like that shown in the left panel of
Fig.~\ref{fig:one}, while the orbit on the right tends to a box orbit like
that in the right panel of Fig.~\ref{fig:one}.}\label{fig:two}
\end{figure}

In the central section of the axis, $\Jz$, which quantifies the height of a
box orbit, largely quantifies $v_z$, while $\Jr$ quantifies $v_x$ and $v_y$.
Outside the central section of the symmetry axis, $\Jr$, which quantifies the
radial excursions of a loop orbit, largely quantifies $v_z$ while $\Jz$
quantifies $v_x$ and $v_y$. In both sections of the symmetry axis, the
distributions of $v_x$ and $v_y$ are guaranteed to be identical because they
are set by the way $\fJ$ depends on the same argument: $\Jr$ in the
central section and $\Jz$ further out.  If we wish to avoid rapidly changing
velocity distributions as we move between the central and outer sections of
the symmetry axis, we should relate the way $f$ depends on $\Jr$ and $\Jz$
along the line in the plane $\Jphi=0$ that has boxes on one side and loops on
the other.

Since the velocity distribution should be a continuous function of position,
the distributions of $v_x$ and $v_y$ should be nearly the same if we move a
small distance along the $x$ axis from the symmetry axis. Once we are off the
symmetry axis, orbits with non-zero $\Jphi$ contribute to the kinematics.
Fig.~\ref{fig:two} shows two orbits with the same small value of $\Jphi$.
The orbit on the left approaches the symmetry axis away from its central
section, while the orbit on the right approaches just this central section.
If we start from the central section of the symmetry axis and move parallel
to the $x$ axis, it will be orbits like that on the right of
Fig.~\ref{fig:two} that contribute to the kinematics. A change to $v_x$
of such an orbit changes only $\Jr$, while a change to $v_y$ changes both
$\Jr$ and $\Jphi$. Unless we restrict the way $f$ depends on $\Jr$ and
$\Jphi$, there is no guarantee that the distributions of $v_x$ and $v_y$
will be similar at our new location and  on the symmetry axis.

If we move parallel to the $x$ axis from a point on the symmetry axis that
lies outside the central section, it will be orbits like that shown on the
left of Fig.~\ref{fig:two} that contribute to the kinematics. A change
to $v_x$ on an orbit of this type only varies $\Jz$ (which controls the amplitude of
oscillations perpendicular to the symmetry axis), while a change to $v_y$
mainly changes $\Jphi$. Hence the dependence of $f$ on $\Jphi$ and $\Jz$
must be restricted if the condition of approximate isotropy just off the
symmetry axis is to be
satisfied.

In summary, these arguments show that 

\begin{itemize}
 \item At $\Jphi=0$ the derivatives of $f$ with respect to $\Jr$ and 
 $\Jz$ should be related along a line in the plane $\Jphi=0$.

 \item At small $\Jz$, the derivatives of $f$ with respect to $\Jr$ and
 $\Jphi$ should be related in the limit $\Jphi\to0$.

 \item At larger $\Jz$, the derivatives of $f$ with respect to $\Jz$ and
 $\Jphi$ should be related in the limit $\Jphi\to0$.
\end{itemize}

In Section~\ref{sec:example} we will show that a DF that does not guarantee
near $xy$ isotropy all along the minor axis, generates unphysical density
distributions.

\subsection{Essential restrictions}\label{sec:restrict}

Velocity isotropy would be guaranteed if
the DF were a function $f(H)$ of the Hamiltonian. Then the derivatives of $f$
with respect to $\Jr$ and $\Jphi$ would be in the ratio of orbital
frequencies
\[
{\p f/\p \Jphi\over\DD f/\DD \Jr}={\Omega_\phi\over\Omega_r}.
\]
For definiteness, we restrict ourselves to cored models.
Orbits that are confined to the  core will be
essentially harmonic, with the consequence that for these orbits
$\Omega_\phi/\Omega_r\simeq\frac12$. We conclude that we can ensure that the
velocity distribution parallel to the equatorial plane tends smoothly to the mandatory
central isotropy by requiring that
\[
\lim_{|\bJ|\to0}{\DD f/\DD \Jphi\over\DD f/\DD \Jr}=\fracj12.
\]
It is not hard to see that satisfaction of the very similar condition
\[\label{eq:JpJr}
\lim_{\Jphi\to0}{\DD f/\DD \Jphi\over\DD f/\DD\Jr}=\fracj12\hbox{ at small }\Jz
\]
ensures that the velocity distribution in the $xy$ plane tends smoothly
to isotropy as one approaches any point on the central section of the
symmetry axis.

Points on the symmetry axis and outside the central section are reached by
orbits with $\Jphi=0$ but significantly non-zero $\Jz$.
In this region, $\Jr$ quantifies the vertical velocity component, which is
unrestricted, while $\Jphi$ and $\Jz$ quantify the two tangential components
of velocity, which should have nearly identical distributions. By the same
chain of argument we deployed above, we infer that the condition of
approximate isotropy in $v_x$ and $v_y$ will be satisfied if
\[
\lim_{\Jphi\to0}{\DD f/\DD \Jphi\over\DD f/\DD\Jz}=\lim_{\Jphi\to0}{\Omega_{\phi}\over\Omega_z}.
\] 
The limiting frequency ratio required here is unity, as one may convince
oneself in two ways: (i) use the Torus Mapper \citep{JJBPJM16} to compute the
frequencies of orbits for diminishing $|\Jphi|$; or (ii) recall that
$\Omega_\phi-\Omega_z$ is the frequency at which the orbital
plane precesses, and that symmetry requires this frequency to be zero for
an orbit that passes right over the pole of the potential. Thus we conclude
that velocity isotropy near the symmetry axis requires
\[\label{eq:JpJz}
\lim_{\Jphi\to0}{\DD f/\DD \Jphi\over\DD f/\DD\Jz}=1.
\] 
  
\subsection{Implementing the restrictions}

\cite{Posti2015} describe a general procedure for constructing DFs that
depend on the actions through a function
\begin{equation}\label{for:kj}
 \kJ = \Jr + \etaphi|\Jphi| + \etaz\Jz,
\end{equation}
that is linear and homogeneous in $\bJ$. Without loss of generality, the
coefficient of $\Jr$ can be taken to be one (B14). \cite{Posti2015} confined their
attention to the case $\etaphi=\etaz$ in which $k$ and therefore $f$ become
functions of $(\Jr,L)$. The model that $f$ then generates is spherical.
B14 had earlier shown in the special  case of the isochrone
\citep{Henon1960} that if one chooses $\etaz>\etaphi$, the model generated by
$f$ is flattened. This idea was later exploited by \cite{DasBinney2016} to
model the flattened inner stellar halo of our Galaxy.

Since for a DF $f[\kJ]$
\[
{\p f/\p J_i\over\DD f/\DD J_j}
={\p k/\p J_i\over\DD k/\DD J_j}={\eta_i\over\eta_j},
\]
adopting constant values of $\etaphi$ and $\etaz$ is not consistent with
the restrictions derived in Section~\ref{sec:restrict}. Hence we replace the
coefficient of $\Jphi$ in $\kJ$ by a function $\Hphi$:
\begin{equation}\label{for:hphi}
 \kJ = \Jr + \Hphi(\bJ,\etaphi,\etaz)|\Jphi| + \etaz\Jz.
\end{equation}
We require $\Hphi$ to be a continuous function of $\bJ$ such
that
\begin{equation}\label{for:con1}
 \Hphi(\vJ,\etaphi,\etaz)\to
\begin{cases}
\etaz & \Jphi/|\vJ|\to0 \\
\frac{1}{2} & |\vJ|\to0 \\
\etaphi &\text{otherwise}.
\end{cases}
\end{equation}
We further require
\begin{equation}\label{for:con2}
 \Hphi(\bJ,\eta,\eta) = \eta \ \ \  \forall\ \bJ
\end{equation}
so that, when $\eta\equiv\etaphi=\etaz$, $k(\bJ)$ depends on $\Jphi$ and
$\Jz$ through the total angular momentum $L=\Jz+|\Jphi|$ and the generated
model is spherical.

We satisfy the conditions~(\ref{for:con1}) and (\ref{for:con2}) by writing
\begin{equation}\label{for:hphi2}
 \Hphi(\bJ,\etaphi,\etaz) = \mIJ\etaphi + (1-\mIJ)\bigl[\mIIJ\etaz +
 (1-\mIIJ)\eta_0\bigr],
\end{equation}
where $\mIJ$ is a smooth function such that 
\begin{equation}\label{for:m1}
 \mIJ(\Jphi)\to
 \begin{cases}
  0 & |\Jphi|\to0 \\ 1 & \text{otherwise}, 
 \end{cases}
\end{equation}
and $\mIIJ$ is a smooth function such that 

\begin{equation}\label{for:m2}
 \mIIJ(\Jphi,\Jz)\to
 \begin{cases}
  0 & (\Jphi,\Jz)\to(0,0) \\ 1 & \text{otherwise}.
 \end{cases}
\end{equation}
Finally, we require the function $\eta_0(\etaphi,\etaz)$ in equation
(\ref{for:hphi2}), to which $\xi$ tends as $(\Jphi,\Jz)\to(0,0)$, to satisfy

\begin{equation}\label{for:con3}
 \eta_0(\etaphi,\etaz)\to 
 \begin{cases}
  \eta & \text{if } \etaphi=\etaz\equiv\eta \\
  \frac{1}{2} & \text{if } \etaz>\etaphi.
 \end{cases}
\end{equation}

\begin{figure}
 \centering
 \includegraphics[width=.8\hsize]{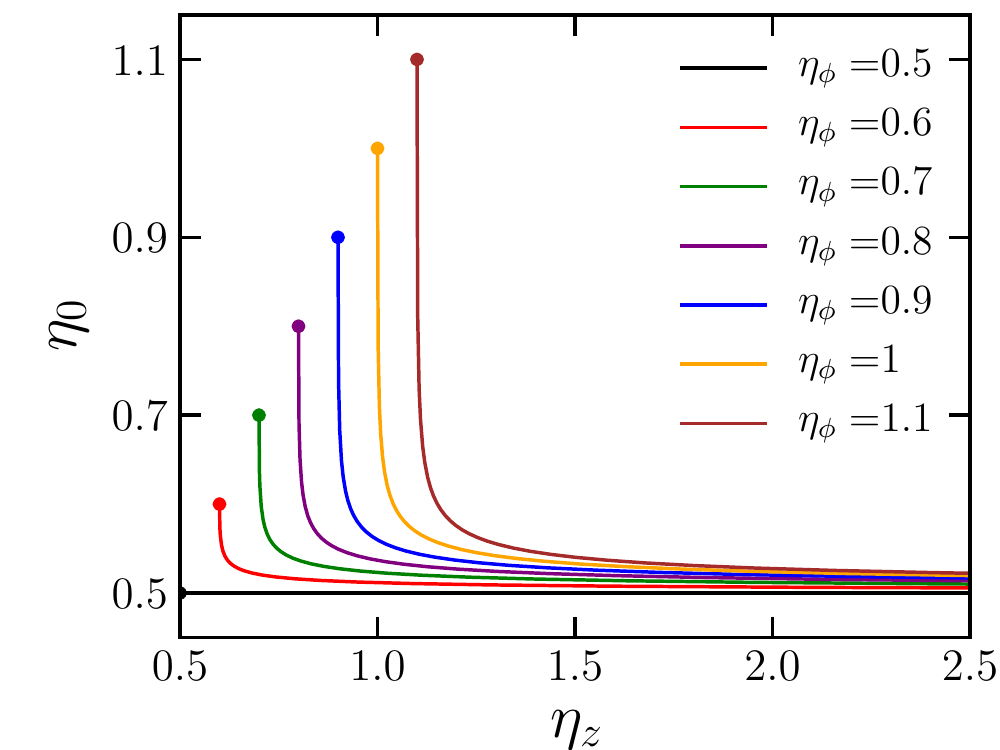}
 \caption{$\eta_0$ (equation \ref{for:ups}) as function of $\etaz$,
 given different $\etaphi$.}\label{fig:ups}
\end{figure}

Functions that satisfy conditions (\ref{for:m1}) to (\ref{for:con3}) are
\begin{align}\label{for:ups}
 \mIJ(\Jphi) &= \frac{\Jphi^2}{\Jf^2 + \Jphi^2},\cr
 \mIIJ(\Jphi, \Jz) &= \frac{\Jphi^2 + \Jz^2}{\Jf^2 + \Jphi^2 + \Jz^2},\cr
 \eta_0(\etaphi,\etaz) &= \etaphi\frac{1 + 10\sqrt{\etaz-\etaphi}}{1 +
 20\etaphi\sqrt{\etaz - \etaphi}},
\end{align}
where $J_0$ is a scale action characteristic of the system's core.  The form
of $\eta_0$ specified by equations (\ref{for:ups}) satisfies  the
conditions (\ref{for:con3}): (i) it ensures a continuous transition between
flattened models ($\etaz>\etaphi$) and spherical models
($\etaz=\etaphi\equiv\eta$); (ii) in the case of even small flattening, it
quickly tends to $\frac12$.
Fig.~\ref{fig:ups} plots $\eta_0$ as a function of $\etaz$ for given
values of $\etaphi$. Substituting for $\mIJ$ and $\mIIJ$ from equations
(\ref{for:ups}) allows one to rearrange equation (\ref{for:hphi2}) to
\begin{equation}\label{for:newB14}
\Hphi = \frac{1}{1 + \tJphi^2}\Biggl(\etaphi\tJphi^2 + \frac{\eta_0 + 
\etaz(\tJphi^2 + \tJz^2)}{1 + \tJphi^2 + \tJz^2}\Biggr), 
\end{equation}
where  $\tJphi\equiv |\Jphi|/\Jf$ and $\tJz\equiv\Jz/\Jf$ are dimensionless
actions.

\begin{figure}
 \centering
 \includegraphics[width=.8\hsize]{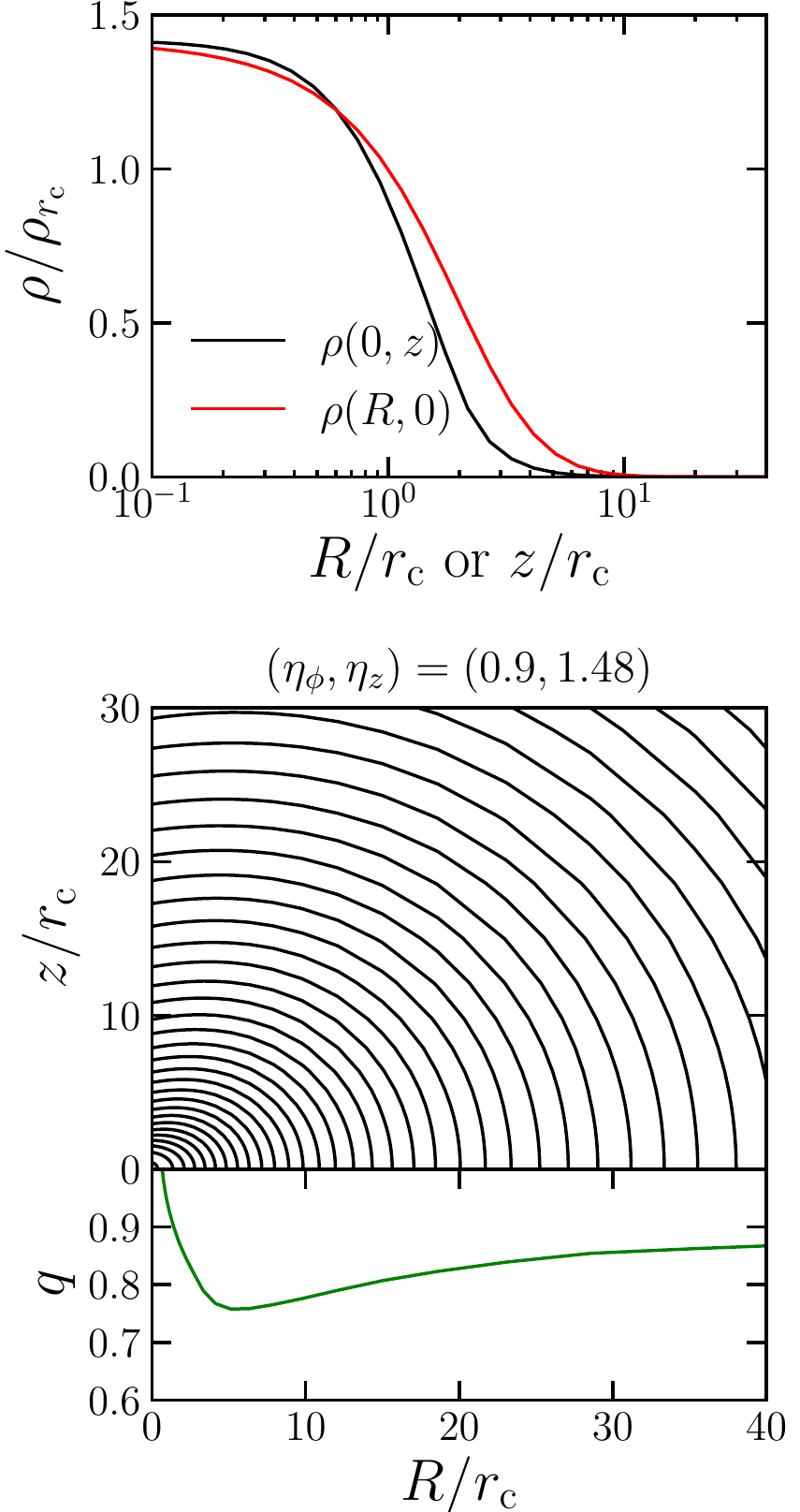}
 \caption{Top panel: major (red) and minor (black) axis density profiles of a
 model in which $\kphieta$ (equation~\ref{for:kphi}). The profiles are normalised 
 to the density on the major axis at the core radius $\rc$, $\rho_{\rc}\equiv
 \rho(\rc,0)$. Middle panel: iso-density contours in the meridional plane. Bottom 
 panel: profile of the axis ratio $q=c/a$, which is obtained by fitting
 ellipses with semi-axis lengths $a,c$ to iso-density contours. The model's
 parameters are $(\alpha,\etaphi,\etaz)=(1,0.9,1.48)$.}\label{fig:flmod}
\end{figure}

\begin{figure*}
 \centering
 \includegraphics[width=.9\hsize]{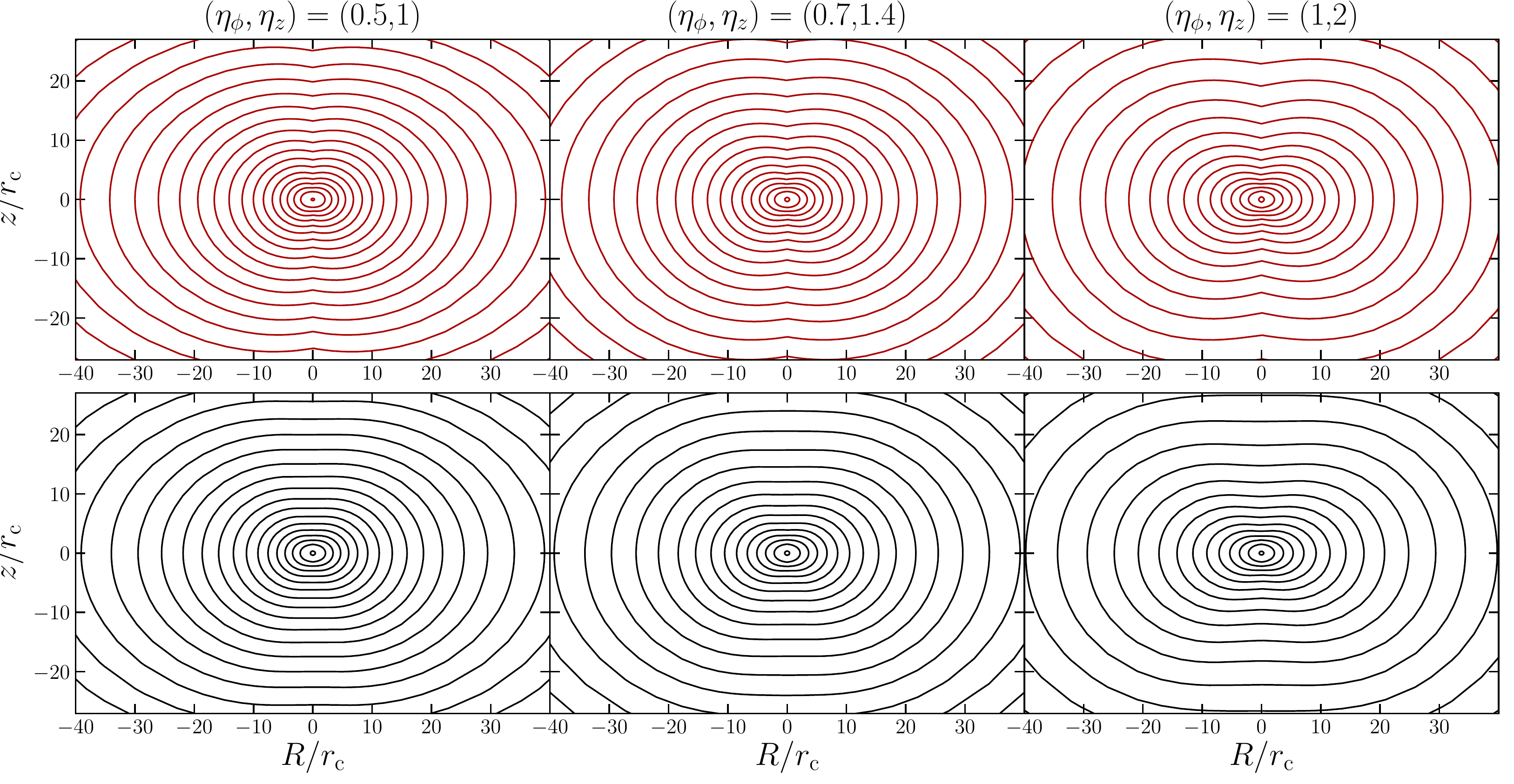}
 \caption{Upper row: iso-density maps of three flattened models obtained from
 the DF (\ref{for:PasDF}) with $\kphieta$ in equation (\ref{for:kphi}). 
 Lower row: same as the top panels, but with $\kphixi$ in equation 
 (\ref{for:kphi}). All models have $\alpha=1$, while from left to right 
 the models have ($\etaphi, \etaz$) = (0.5,1), (0.7,1.4) and (1,2).}\label{fig:newmod}
\end{figure*}

\begin{figure*}
 \centering
 \includegraphics[width=.8\hsize]{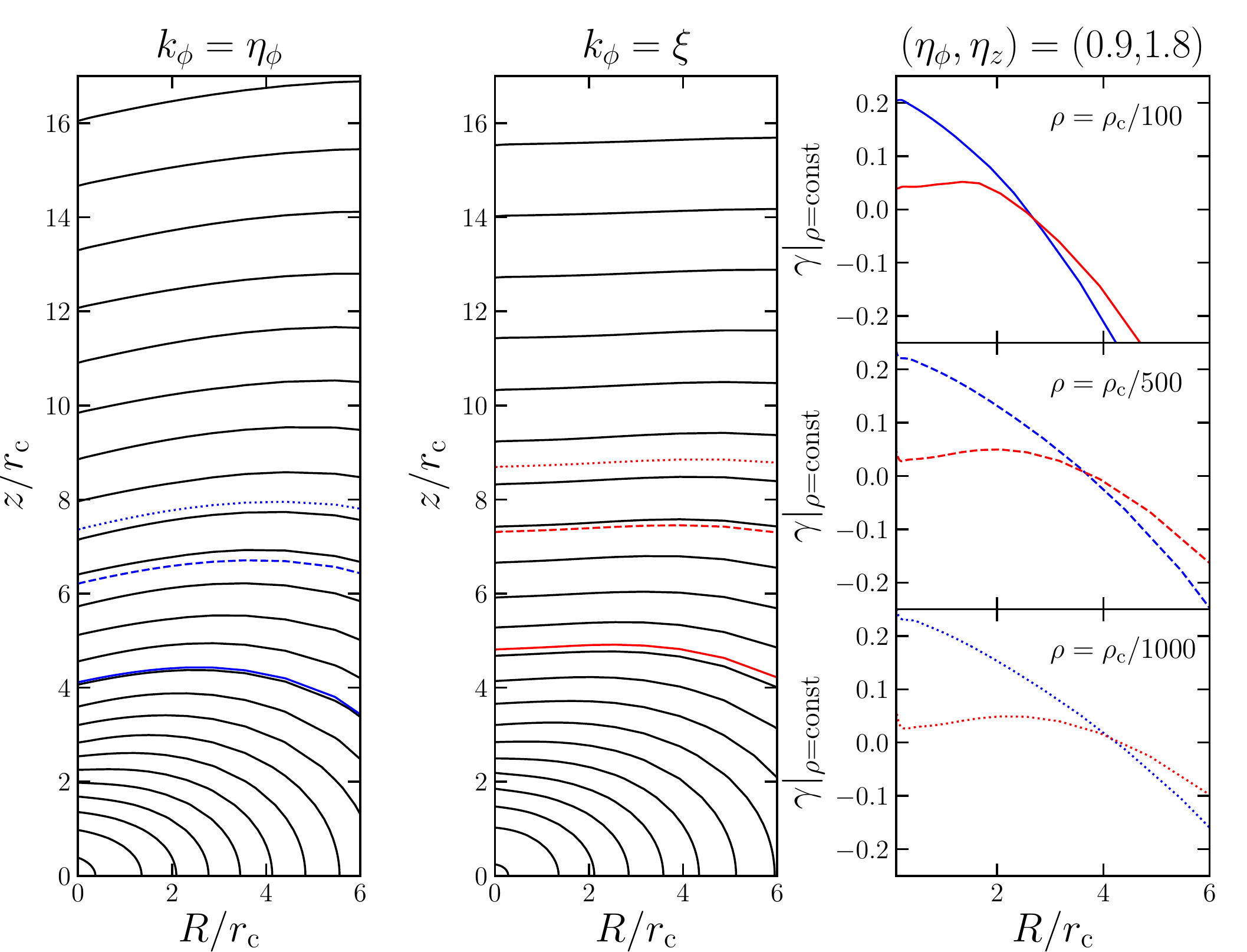}
 \caption{Left panel: iso-density contours near the symmetry axis of a model
 with $(\alpha,\etaphi,\etaz) = (1,0.9,1.8)$ generated using $\kphieta$. Middle panel:
 same as the left panel, but for a model generated using $\kphixi$.
 Right panel: the gradient of isodensity surfaces (\ref{for:logs}) as a function
 of $R$ at three different values of $\rho$, namely from top to bottom $\rho=0.01, 
 \rhoc,0.002\rhoc$ and $\rho=0.001\rhoc$, with $\rhoc$ equal to the model central
 density. Data for $\kphieta$ are plotted in blue and those for $\kphixi$ in 
 red.}\label{fig:dp}
\end{figure*}

\section{Worked examples}\label{sec:example}

\begin{figure*}
 \centering
 \includegraphics[width=1\hsize]{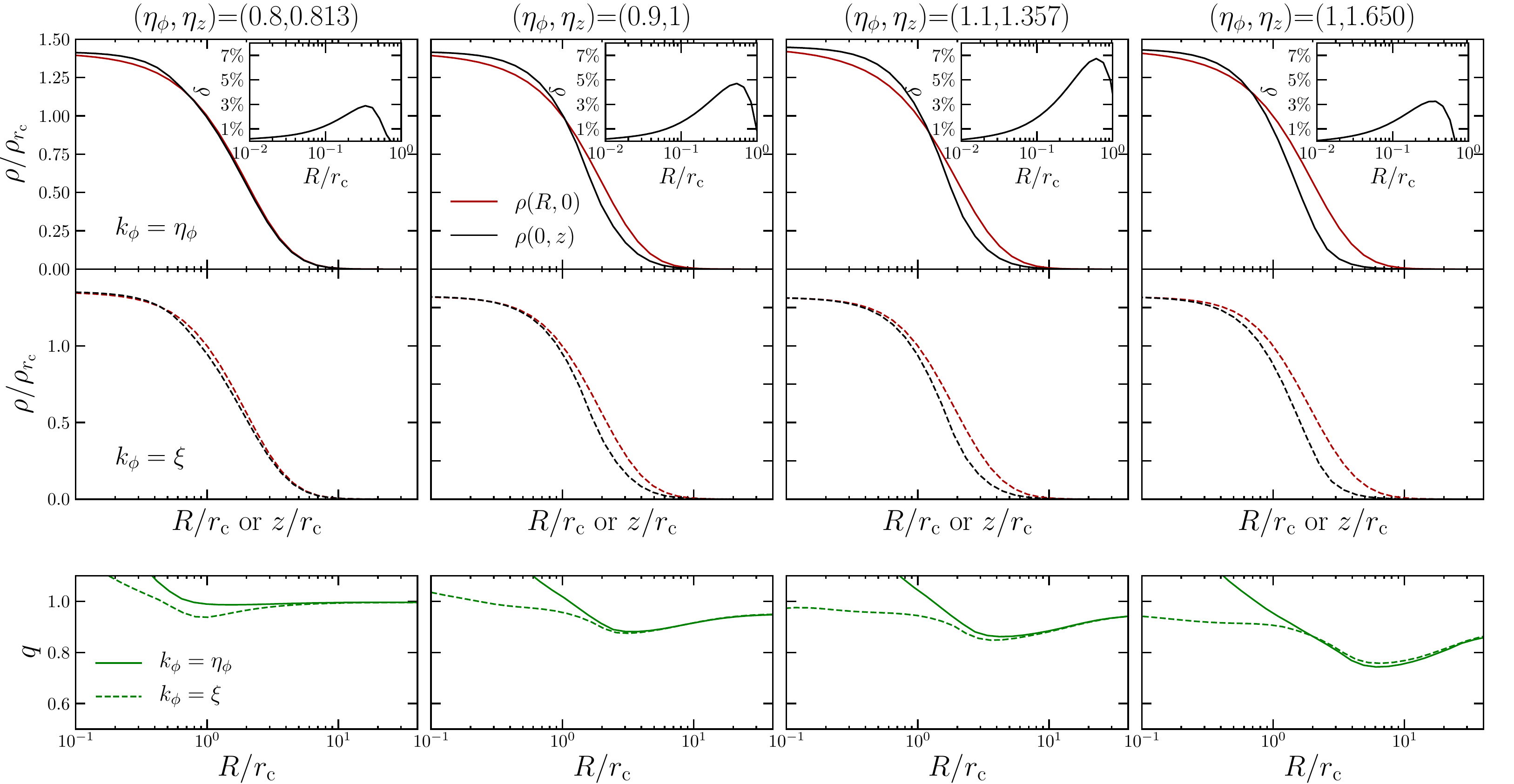}
 \caption{Top and middle rows show density profiles along major (red) and minor
 (black) axes. The top row shows results obtained when $\kphieta$ in equation 
 (\ref{for:kphi}), while the second row shows corresponding results when 
 $\kphixi$. Bottom row: axis-ratio $q=c/a$ as a function of the semi-major axis length. 
 Solid curves describe the profiles plotted in the top row, while broken curves
 describe the profiles of the second row. The sub-panels in the top row show 
 $\delta$ (equation~\ref{for:delta}) as a function of $R$ for $R\le\rc$.
 From the left to right the models' parameters are ($\etaphi,\etaz) = 
 (0.8,0.813)$, (0.9,1), (1,1.357) and (1,1.650). All models have $\alpha=1$.}\label{fig:prol2}
\end{figure*}

We illustrate the benefit of using $\Hphi$ (equation~\ref{for:newB14}) rather
than $\etaphi$ as the coefficient of $\Jphi$ in the function $\kJ$ by
computing some flattened models derived from the DF that \cite{Pascale2018}
introduced to model the Fornax dwarf spheroidal galaxy -- \cite{Pascale2019}
already explored the range of spherical models that this DF generates. The DF is
\begin{equation}\label{for:PasDF}
 \fJ = f_0\exp\biggl[-\biggl(\frac{\kJ}{\Jf}\biggr)^{\alpha}\biggr],
\end{equation}
where $\alpha$ is a positive, dimensionless constant that primarily controls
the model's density profile, $\Jf$ is a scale 
action that sets the size of the model's core, and $f_0$ is a normalising
constant that ensures that $(2\pi)^{-3}=\int\d^3\bJ\,\fJ$. Here we assume $\alpha=1$. 
In the interests of generality, in equation (\ref{for:PasDF}) we define 
\begin{equation}\label{for:kphi}
 \kJ = \Jr + \kphi|\Jphi| + \etaz\Jz,
\end{equation}
where 
\begin{equation}
 \kphi = 
 \begin{cases}
  \etaphi & 			\text{old models (equation \ref{for:kj})} \\
  \xi(\bJ,\etaphi,\etaz) & 	\text{new models (equation \ref{for:newB14})}.
 \end{cases}
\end{equation}
We will refer to the old models, with constant $\etaphi$ as coefficient 
of $\Jphi$ in equation (\ref{for:kj}), as $\kphieta$, and to the new models, 
with $\xi$ as coefficient of $\Jphi$ (equation \ref{for:newB14}), as $\kphixi$.

Fig.~\ref{fig:flmod} plots various quantities for a model computed with
$\kphieta=0.9$. The top panel shows the density profiles along the model's 
major (red) and minor (black) axes, with distances scaled to the core 
radius $\rc$, defined as the distance down the major axis at which 
\begin{equation}\label{for:gamma}
\frac{\DD\ln\rho}{\DD\ln R}\biggr|_{(R=\rc,z=0)} =
-\frac{1}{2}.
\end{equation}
The middle panel of Fig.~\ref{fig:flmod} shows 
isodensity contours in the meridional plane, and the bottom panel shows
the axis ratios of these contours that one obtains by fitting ellipses 
to the contours as explained in Appendix~\ref{app:flat}. 
These plots reveal unphysical features that derive from the use of $\kphieta$
in equation (\ref{for:kphi}).

At the left edge of the middle panel, the contours are sloping down to the
left, reflecting a depression in the density along the minor axis, and, worse
still, a discontinuity in the direction of the normals to isodensity surfaces
where they cut the symmetry axis. The bottom panel shows a rapid decrease in
the model's flattening as one approaches the centre, which sends the axis
ratio $q\equiv c/a$ through unity to values indicative of prolateness before the
centre is reached. This feature can also be seen in the top panel, where,
when approaching the center, the density along the $z$-axis becomes larger 
than the density computed along the $R$-axis of the meridional plane.
These unphysical features arise from the failure of the DF (\ref{for:PasDF})
with $\kphieta$ rather than $\kphixi$ (equation~\ref{for:kphi}) to respect 
restriction on the velocity distribution along the minor axis that we
derived in Section~\ref{sec:restrict}.

Fig.~\ref{fig:newmod} compares six models computed with different
$\kphi$: in the models shown in the upper row $\kphieta$, while in the
models of the lower row $\kphixi$. The models have similar flattenings but 
their radial bias increases from left to right: their parameters are 
$(\etaphi, \etaz)$ = (0.5,1), (0.7,1.4) and (1,2) and flattening increases 
with the ratio $\etaz/\etaphi$, while increasing both $\etaphi$ and $\etaz$
increases the radial bias of the velocity distribution. All models have
$\alpha=1$.

The isodensity contours plotted in the upper panels of Fig.~\ref{fig:newmod} have 
clearly discontinuous slopes across the minor axis. The contours plotted in
the lower panels do not show this unphysical discontinuity, so using
$\kphixi$ rather than $\kphieta$ banishes cuspy isodensity surfaces. 
The left and middle panels of Fig.~\ref{fig:dp} show the iso-density 
contours along the minor axis on an enlarged scale for models with
$(\etaphi,\etaz)=(0.9,1.8)$. The right column in this figure shows the
variation with $z$ of the gradients of contours  
\begin{equation}\label{for:logs}
 \gamma(R,z) \equiv \left(\frac{\DD z}{\DD R}\right)_{\rho={\rm const}},
\end{equation}
at three values of $\rho$: $\rho=0.01\rhoc,0.002\rhoc,0.001\rhoc$, with $\rhoc$ 
equal to the model central density. Positive values of $\gamma$ indicate a
depression along the reference isodensity contour when approaching 
the minor axis. A depression is not unphysical -- `peanut' bulges of disc 
galaxies have such depressions -- but a non-zero value of $\gamma$ as $R\to0$ 
{\it is} unphysical. Blue curves show $\gamma(R)$ when $\kphieta$, while
red curves show $\gamma(R)$ when $\kphixi$. We see that use of $\Hphi$ 
rather than $\etaphi$ ensures that the slopes of isodensity contours are 
only slightly positive near the axis and vanish on the minor axis.

\begin{figure}
 \centering
 \includegraphics[width=.8\hsize]{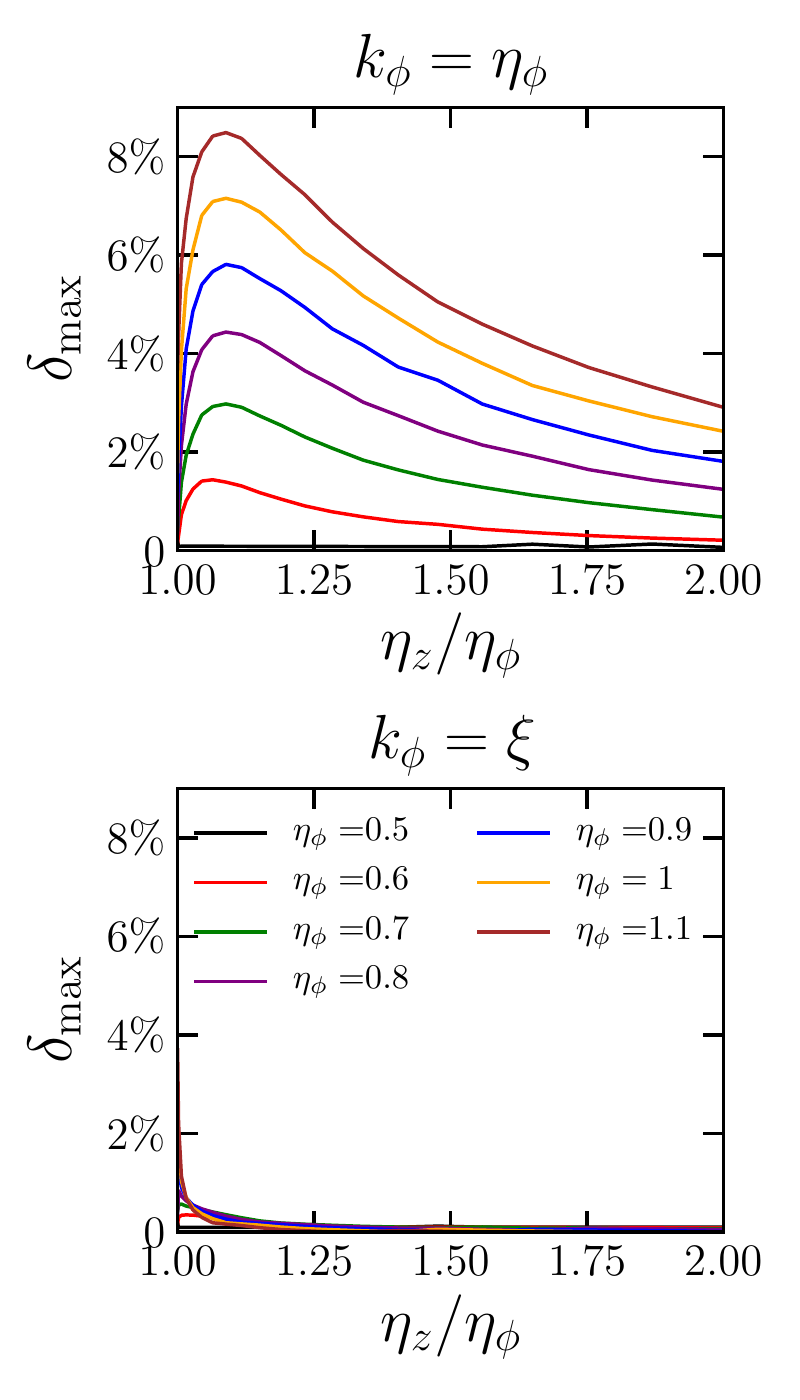}
 \caption{The maximum value $\delta_{\rm max}$ of $\delta$ 
 (equation~\ref{for:delta}) as a function of the ratio $\etaz/\etaphi$ 
 for models with different $\etaphi$ Upper panel: when $\kphieta$; 
 lower panel: when $\kphixi$. All models have $\alpha=1$.}\label{fig:prol}
\end{figure}

Fig~\ref{fig:prol2} shows density profiles and axis ratios 
for models with flattenings that increase from left to right. Panels in the
top row show density profiles along the major (red) and minor (black) axes 
for models computed with $\kphieta$, while below them we show the corresponding
models with $\kphixi$. In the top row, the red curves fall below the black
curves at $R\la\rc$, implying that these generally oblate models have 
prolate cores. In the panels of the middle row, the red and black curves 
approach $R=0$ together. The bottom row shows plots of the axis ratios as a
function of semi-major axis. The models with $\kphieta$ have central axis 
ratios significantly greater than unity, whereas the more flattened
models with $\kphixi$ have axis ratios that are always less than
unity. In the least flattened of the models that uses $\kphixi$ (extreme
left panel), $c/a$ does exceed unity at $R<0.3\rc$, but this model is so nearly
spherical that this tendency to central prolateness is of little
significance. The model just to its right becomes marginally prolate at
$R<0.2\rc$, where the density is extremely close to the central density and
even a tiny angular variation in density generates a significant value of
$1-|c/a|$.

Further exploration of any tendency to prolateness in the core is facilitated
by defining the diagnostic
\begin{equation}\label{for:delta}
 \delta(u) \equiv 1 - \frac{\rho(u,0)}{\rho(0,u)}.
\end{equation}
The small panels of Fig.~\ref{fig:prol2} are plots of $\delta(u)$ in models
in which $\kphieta$. Clearly, in an oblate model $\delta$ should be negative,
but the figure shows that $\delta>0$ in the core to an extent that varies with
($\etaphi, \etaz$). Fig.~\ref{fig:prol} plots for several values of $\etaphi$ 
the peak value of $\delta$ as a function of the ratio $\etaz/\etaphi$ that
controls a model's flattening.  The top panel is for models that use $\kphieta$
and the bottom panel is for models that use $\kphixi$. In the top panel, the
largest values of $\delta$, and therefore the most prolate cores, occur in models
with $\etaz/\etaphi\simeq1.1$ i.e., nearly spherical models, as is to be expected, 
and in models with the largest values of $\etaphi$ and therefore the greatest
radial velocity bias.  In the most radially biased model ($\etaphi=1.1$), the
peak in $\delta$ reaches $0.08$ and this model is prolate throughout the core. 
In the bottom panel, $\delta<0.01$ even in the most radially biased and least
flattened model. In the vast majority of models it is much smaller. Thus
replacing $\etaphi$ with $\Hphi$ essentially resolves the issue of prolate 
cores in addition to banishing cusped isodensity contours.
 
\begin{figure*}
\centering
\includegraphics[width=.9\hsize]{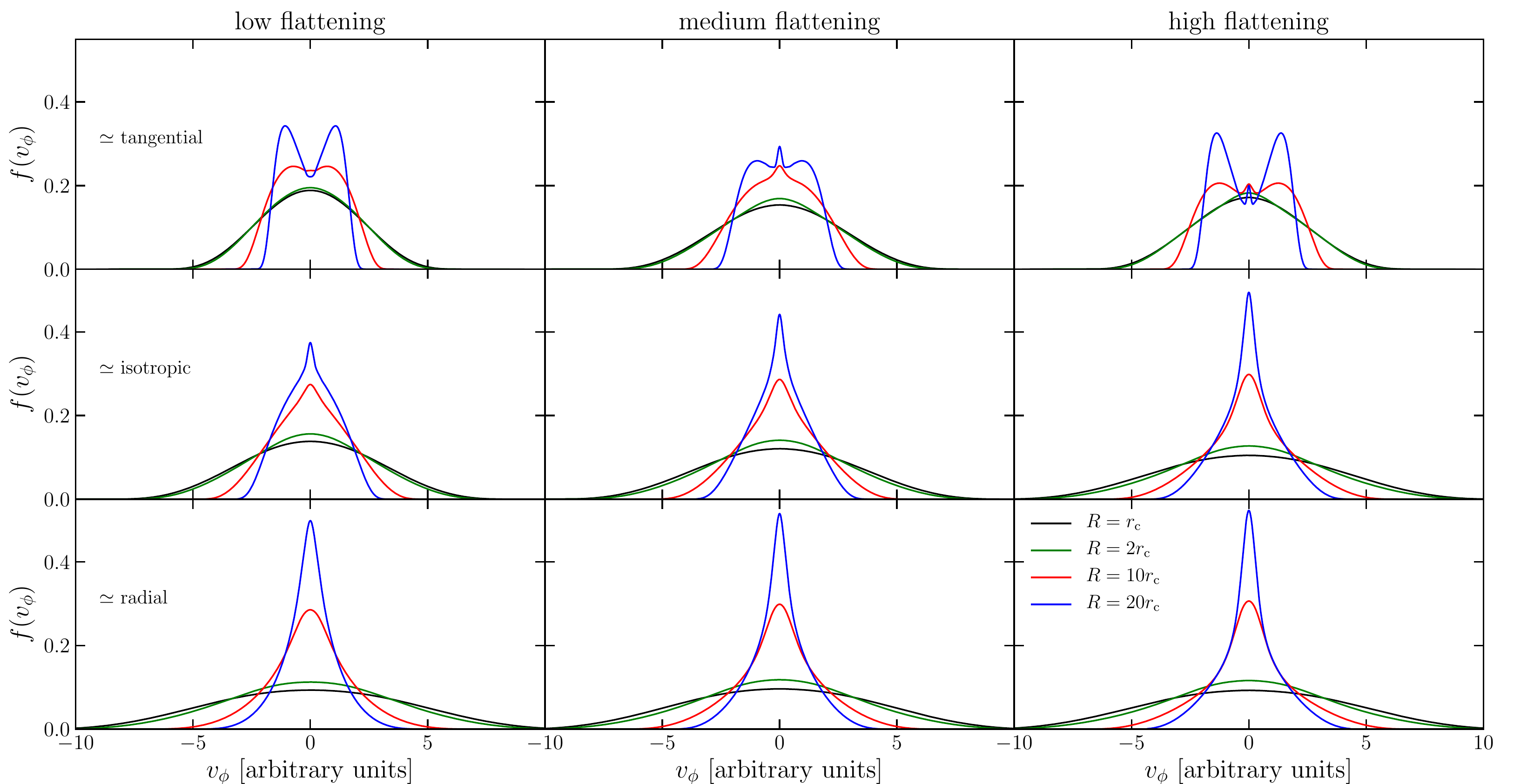}
\caption{Distributions of $v_{\phi}$ at six locations in nine models with 
$\kphixi$ that differ in their flattening (by column) and by their degree of 
radial bias (by row). The distributions for $R<\rc$ are not shown because they 
are very similar to those for $R=\rc$. From left to right, and from top to bottom,
the models have ($\etaphi,\etaz$) = (0.5,0.6), (0.6,0.9), (0.5,1),
(0.7,0.9), (0.8,1.1), (0.9,1.5), (1.1,1.2), (1,1.5) and (1,2). All models have 
$\alpha=1$.}\label{fig:vt}
\end{figure*}

\subsection{Distribution of $v_{\phi}$}

\cite{PifflPenoyreB} encountered a problem with the distribution of azimuthal
velocity components that occurs already in spherical models of the B14 type:
when the model has a radial bias, a cusp around $v_\phi=0$ appears in the
$v_\phi$ distribution because 
\begin{equation}
  \lim_{v_\phi\to0^+}{\p\over\p v_\phi}\int\d v_R\d v_z\,f =  \\ 
  -\lim_{v_\phi\to0^-}{\p\over\p v_\phi}\int\d v_R\d v_z\,f<0.
\end{equation}
When the model has tangential bias, the cusp is replaced by a dimple, so the
above limit is positive. To eliminate this problem, they forced the dependence
of $f$ on the actions in the limit $\Jphi\to0$ to mirror that of the
Hamiltonian. To achieve this goal with $k$ defined by equation
(\ref{for:hphi}), we would have to make $\etaz$ depend on the actions such
that it tends to $\Omega_\phi/\Omega_r$ as $\Jphi\to0$. Fig.~\ref{fig:vt}
explores the consequences of our failure to take this step by
showing $v_\phi$ distributions at several locations in nine models of varying
flattening and radial bias. The approximately isotropic and radially biased
models have unexceptionable $v_\phi$ distributions. At $R\ga10\rc$ the
$v_\phi$ distributions of the tangentially biased models do have anomalous
central shapes. The double humped nature of these profiles simply reflects 
the absence of a part of the DF that is odd in $\Jphi$; such a component
would reinforce one hump at the expense of the other, yielding the skew
$v_\phi$ distribution of a rotating model. The anomalous features that one
would ideally eliminate by the method of \cite{PifflPenoyreB} are the 
central spikes displayed by two of the tangentially biased models. 

The difference between the anomalous features in these $v_\phi$ distributions
and those presented by \cite{PifflPenoyreB} probably arises because the latter
were adapting the DFs of \cite{Posti2015}, which generate models with central
density cusps, rather than building cored models.

\section{Discussion}\label{sec:discuss}

Action space is simply a way of cataloguing orbits, and one may ask why one
is not at liberty to populate those orbits as one pleases by setting the DF
to an arbitrary, normalisable, non-negative function of $\bJ$. The proper
response to this claim is to say `yes you can populate the orbits as you
please, but there are two reasons why the schemes of population occurring in
real systems are restricted.' 

The first reason is the requirement for self-consistency. It is possible that
certain DFs are not consistent with an essentially integrable self-consistent
potential. For example, some of the potentials encountered during an attempt to
iterate from $\Phi_0$ to a self-consistent $\Phi_n$ might have significant
chaotic zones. Stars would diffuse through these zones, irreversibly
changing $f$.

The second reason why the DFs of real systems will be restricted relates to
the manner in which stars are distributed in phase space.  Imagine stars being
scattered like confetti, or shrapnel from a shell burst at a large number of
locations. Then the initial conditions will be smooth functions of
$\vx$ and of $\vv$ and the resulting DF $\fJ$ will not be one that induces
discontinuities in velocity distributions. Star formation will likewise
create stars with a density that is a continuous function of $(\bx,\bv)$.


\section{Conclusions}\label{sec:conclude}

While any non-negative, normalisable function $\fJ$ specifies a
self-consistent stellar system, the system it generates will have unphysical
features unless the DF satisfies the constraints (\ref{eq:JpJr}) and
(\ref{eq:JpJz}) on its derivatives. These constraints arise from the
requirement that the velocity distribution at points near the symmetry axis
should differ little from the (axially symmetric) velocity distributions on
the symmetry axis.
If the DF is a function only of the Hamiltonian $H$, these constraints are
automatically satisfied. 

We have experimented with an algorithm that generates DFs for flattened and
possibly radially biased models that are consistent with the DF tending to a
function of the Hamiltonian as $\Jphi$ approaches zero. We have shown that
the resulting models are free of the unphysical features near the symmetry
axis that disfigure models based on the simpler DFs proposed by B14. The new
DFs provide a promising basis for modelling different components of globular
clusters, dwarf spheroidal galaxies and galactic bulges.

\bibliography{paper}
\bibliographystyle{mnras}

\appendix
\section{Quantifying flattening}\label{app:flat}

Each iso-density contour is characterised by a collection of $\{\Ri,\zi\}$
points, with $i=1,...,N$, such that $\rho(\Ri,\zi)=\hbox{constant}$. 
For that contour we define
\begin{equation}
 \chi^2=\sum_{i=0}^{N}\biggl[1 - \biggl(\frac{\Ri^2}{a^2} + \frac{\zi^2}{c^2}\biggr)\biggr]^2,
\end{equation}
where $a$ and $c$ are, respectively, the semi-major and the semi-minor 
axes for an oblate model, viceversa for a prolate model. We find the best 
($a,c$) as the solution of
\begin{equation}
 \frac{\DD\chi^2}{\DD a^2} = \frac{\DD\chi^2}{\DD c^2} = 0,
\end{equation}
which is
\begin{equation}\label{for:a}
 a^2 = \frac{(\sumiN\Ri^2\zi^2)^2 - \sumiN\Ri^4\sumiN\zi^4}
 {\sumiN\Ri^2\zi^2\sumiN\zi^2 - \sumiN\Ri^2\sumiN\zi^4},
\end{equation}
and
\begin{equation}\label{for:b}
 c^2 = \frac{(\sumiN\Ri^2\zi^2)^2 - \sumiN\Ri^4\sumiN\zi^4}
 {\sumiN\Ri^2\zi^2\sumiN\Ri^2 - \sumiN\Ri^4\sumiN\zi^2}.
\end{equation}

\end{document}